\documentclass[preprint]{aastex}
\slugcomment{Submitted to ApJ Letters}

\newcommand{\etal}{{\it et al.\/}}

\begin{document}
\title{The Edge of the Solar System}
\author{R. L. Allen\altaffilmark{1}, G. M. Bernstein\altaffilmark{1}}
\affil{Department of Astronomy, University of Michigan,
830 Dennison Bldg., Ann Arbor, MI 48109,
rhiannon, garyb@astro.lsa.umich.edu}

\and
\author{R. Malhotra\altaffilmark{2}}
\affil{Department of Planetary Sciences, University of Arizona,
1629 E.~University Blvd., Tucson, AZ 85721,
renu@lpl.arizona.edu}

\altaffiltext{1}{Visiting Astronomer, National Optical Astronomy
Observatories, which is operated by the Association of Universities
for Research in Astronomy, Inc., under contract to the National
Science Foundation.}
\altaffiltext{2}{Staff Scientist, Lunar and Planetary Institute,
which is operated by the Universities Space Research Association under
contract No.~NASW-4574 with the National Aeronautics and Space
Administration.}

\begin{abstract}
The population of the Kuiper Belt within 50~AU of the Sun has likely
been severely depleted by gravitational perturbations from the giant
planets, particularly Neptune.  The density of Kuiper Belt objects is
expected to be two orders of magnitude higher just beyond 50 AU, where
planetary perturbations are insignificant.  In 1998 and 1999, we
surveyed for Kuiper Belt Objects (KBOs) in 6 fields of the ecliptic
(total sky area 1.5 deg$^2$) to limiting magnitudes between $R=24.9$
and $R=25.9$.  This is deep enough to detect KBOs of diameter $\gtrsim
160$~km at a distance of 65~AU.  We detected 24 objects.  None of
these objects, however, is beyond 53~AU.  Our survey places a 95\% CL
upper limit of $\Sigma < 5\,{\rm deg}^{-2}$ on the surface density of
KBOs larger than $\sim160$~km beyond 55~AU.  This can be compared to
the surface density of $\sim6\,{\rm deg}^{-2}$ of $\ge160$~km KBOs at
distances 30--50~AU determined from this survey and previous shallower
surveys.  The mean volume density of $D>160$~km KBOs in the 55--65~AU
region is, at $>95\%$ confidence, less than the mean density in the
30--50~AU region, and at most \twothirds\ of the mean density from
40--50~AU.  Thus, a substantial density increase beyond 50~AU is
excluded in this model-independent estimate, implying that some
process or event in the history of the Solar System has truncated the
distribution of 160-km planetesimals at $\sim50$~AU.  A dense
primordial disk could be present beyond 50~AU if it contains only
smaller objects, or is sufficiently thin and inclined to have escaped
detection in our 6 survey fields.
\end{abstract}


\section{Introduction}
Our planetary system should not be expected to be bounded by the orbit
of the ninth planet Pluto, as pointed out by \citet{Ed49} and
\citet{Ku51}.  Our current general understanding of the formation of
our solar system is consistent with the expectation that a large
population of small bodies -- leftovers from the primordial
planetesimal disk in the Solar Nebula -- exists at the present time
beyond the orbits of Neptune and Pluto where planetary accretion
timescales exceed the current age of the solar system.  Indirect
evidence for such a population (now known as the {\it Kuiper Belt})
exists in the orbital properties of the short period comets, as has
been suggested by several authors \citep{Fe80,Du88}. Direct
observational evidence was first obtained with the discovery of 1992
QB1 \citep{Je93}, and has subsequently grown with the detection of
more than 300 objects.  It is notable that all but one of the
currently known Kuiper Belt objects (KBOs) lie within 55 AU of the
Sun.

The semimajor axis zone $30<a<50$ AU is dynamically a very complex
region. The primordial Kuiper Belt population here is expected to have
been extensively sculpted and depleted throughout solar system history
[see review by \citet{Ma00}].  Numerical simulations show that most
primordial KBOs on orbits with $a<36$~AU would have been ejected from
the Kuiper Belt in much less than the age of the solar system
\citep{Ho93, Le93, Du95}.  In the $36<a<42$ AU zone, the dynamical
lifetime of low eccentricity, low inclination orbits is comparable to
the age of the solar system, whereas beyond $a\approx42$ AU the
dynamical lifetime is well in excess of the age of the Solar system.
The orbital migration of the giant planets in early solar system
history is expected to have rearranged the primordial orbital
distribution in the $a<50$ AU zone, by sweeping a significant fraction
of KBOs into eccentric orbits at narrow semimajor axis zones at the
locations of mean motion resonances with Neptune \citep{Ma93, Ma95}.
Furthermore, there exists a significant ``scattered'' KBO population,
in very eccentric orbits with perihelia near $\sim 36$ AU; these KBOs
are thought to have been formed in the $36< a< 42$ AU zone and been
gravitationally scattered during a close encounter with Neptune
\citep{Le97}.

The current estimate of the surface density in the inner Kuiper Belt
is approximately two orders of magnitude less than that expected from
a smooth extrapolation of the surface density from the planetary
region [{\it cf.} \citet{We77}].  It is also approximately two orders
of magnitude less than that required for the formation of the largest
observed KBOs within $10^8$~yrs, before perturbations from the giant
planets curtailed their growth \citep{S96, SCb, KL99}.  These
arguments suggest severe depletion of the Kuiper Belt within 50 AU.

Beyond 50 AU, the gravitational influence of Neptune and the giant
planets is insignificant.  A higher present-day density of objects,
reflective of the primordial surface density of solids in the Solar
Nebula, would be expected in this unperturbed distant Kuiper Belt.
Previous surveys have detected no objects beyond 50~AU, leading
several authors \citep{Do97, Ch99, JLT} to suggest that this
higher-density outer region is in fact absent, and that the primordial
planetesimal disk must have had a cutoff near 50~AU. \citet{Gl98}
calculate the fraction of detected KBOs expected to be beyond 50~AU;
applying their formula to the number of KBOs known today also
indicates that there is a deficiency of distant objects.  These
conclusions, however, depend quite sensitively upon the exponents of
assumed power laws for size and radial distributions of the
population.

In this Letter we report the results of a survey for faint KBOs
covering more than one square degree of the ecliptic.  Most of the
known KBO population lie at $\sim40$~AU and have magnitudes $R\le24$.
At our limiting magnitudes of $R\ge25$, we can detect physically
similar objects to a distance of 65~AU or greater.  We compare the
results of our faint survey with the detected population of similarly
sized KBOs at smaller distances, and test whether the population
density increases beyond 50 AU.  This comparison of the population of
$\gtrsim 160$~km objects is independent of any models or assumptions
for the size distribution of the KBOs in the two regions.

\section{Search for Distant Objects}

Observations were taken using the BTC CCD Mosaic Camera (Wittman
\etal\ 1998) at the Cerro Tololo Interamerican Observatory on 3 nights
in May 1998 and 4 nights in May 1999.  The BTC images 0.25~deg$^2$ of
(non-contiguous) sky per exposure.  Each field was observed for a series
of 20 to 30 eight-minute exposures spread across two consecutive nights. 
These images are registered and summed (with sigma-clipping rejection)
to yield a deep template image of the fixed field.  The template is then
subtracted from each individual exposure after applying an algorithm
to match the point spread functions, thus efficiently removing all
non-moving objects from the images.  These images are first searched for
bright slow-moving objects.  Then each field's images are summed with
displacements to track a potential KBO motion, and the sum image searched
for faint objects.  This last step is repeated several thousand times to
permit detection of KBOs in any bound orbit at distances 30--80~AU.
The entire search procedure is repeated on another pair of nights six to
eight days later.  Candidate KBOs were confirmed using data taken
six to eight days apart.

A detailed description of our processing and detection methods will be 
published at a later date.

Effective areas and limiting magnitudes are determined by inserting
several thousand artificial KBOs into the raw data at a variety of
velocities and positions. We then search for these using the same
methods used to detect real KBOs.  The total area subtended by the six
search fields is $1.5$~deg$^2$, but bright stars, CCD defects, etc.,
limit the {\em effective area} of the search to $1.3$~deg$^2$ for
bright KBOs.  The effective search area drops at fainter magnitudes
due to noise.

The May 1998 observations used the Kron-Cousins $R$ filter, and we
find that the effective area of these three fields of the search is
$0.51\, {\rm deg}^2$ at bright magnitudes, dropping by 50\% at
$R=24.9$ to $R=25.4$. The May 1999 observations, using a wide-band
$VR$ filter \citep{JLC}, reach 50\% completeness for 25.3--25.9 $R$
mag KBOs over effective area of $0.77\,{\rm deg}^2$.  $VR$ magnitudes
are
converted to $R$ assuming a KBO color index of $V-R=0.3$.  Table~1
lists the fields observed for this survey and the total integration
times, limiting magnitudes and effective areas for each.  Several
times more area is covered than for the largest previously published
survey to this depth \citep{LJ98}.

\begin{figure}
\plotone{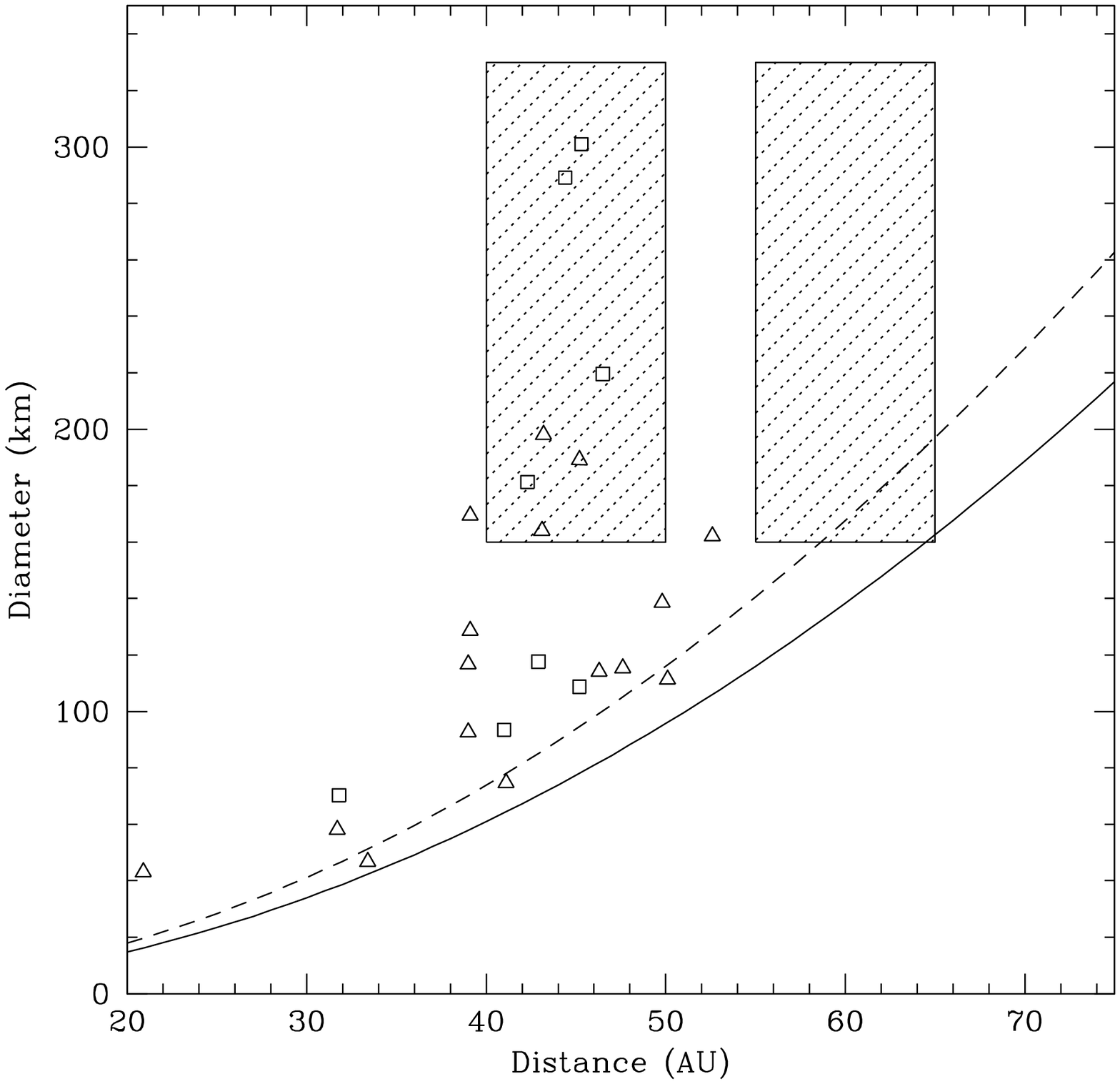}
\caption{
Diameter vs distance for the KBOs discovered in this survey.  Squares
are objects discovered in 1998, triangles are from 1999.  The two
curves denote lines of constant magnitude, corresponding to the 50\% 
completeness limits of the overall and faintest survey fields.  
The two rectangular boxes contain denote
the ``inner'' and ``outer'' regions of the Kuiper Belt described in
the text.  The absence of any objects in the outer sample region
limits the volume density of $\ge160$~km diameter objects at 55--65~AU
to be below that at 40--50 AU at 95\% CL.
\label{dvsa}
}
\end{figure}

Twenty-three KBOs and one Centaur were discovered in the survey.  The
observing scheme produces a 10-day arc for most of these objects.
Orbital parameters and their uncertainties are determined using the
software described in \citet{Be00}.  Although the full orbital
parameters are poorly constrained by such a short arc, the distance is
determined fairly accurately.  This is because the object's apparent
acceleration is due almost entirely to the reflex of the Earth's
transverse acceleration, and hence the apparent acceleration is
inversely proportional to the distance.  Note that this technique
requires that we {\it not} observe at opposition.  Recovery
observations have been made for 7 of the objects, and in all cases the
refined orbits produce distances consistent with the 10-day estimates.
A few of the objects were detected near the edges of the frame and
were outside the field of view for the second (confirming) pair of
nights.  Distances to these objects with only 24-hour arcs are less
certain, and dependent upon the assumption that the orbit is bound.
Table~2 contains a list of detected objects and their heliocentric
distances, and Figure~\ref{dvsa} plots their size vs.~distance under
the assumption of a 4\% albedo.  Distances beyond 53~AU are ruled out
for all of these objects.  Thus, over $1.3\,{\rm deg}^2$, we have zero
detections of objects beyond 53 AU, even though objects with diameters
larger than 160~km would be visible to 65~AU over much of the field.

The diameter (D [km]) of a Kuiper Belt Object can be estimated using
the formula for a uniformly scattering sphere \citep{Ru16}:
\begin{equation}
p D^2 \phi = 9\times10^{16}\,{\rm km}^2 {{R^2 \Delta^2} \over {1\,{\rm AU}^4}} 10^{0.4(m_\odot-m_R)}
\end{equation}
where $p$ is the geometric albedo, taken to be $p\approx0.04$
following \citet{JLT}.  $R$ and $\Delta$ are the heliocentric and
geocentric distances in AU, which are approximately equal for these
large distances; $m_\odot$ and $m_R$ are the apparent magnitude of the
Sun and KBO, respectively; and $\phi$ is a phase correction, which we
ignore for these distant KBOs since the phase angle is always small.

A quantitative characterization of the upper limit on large KBOs in
the 55--65~AU range is as follows: At 55~AU, a magnitude of $R=25.1$
is sufficient to detect $D>160$~km objects, while at 65~AU, the
necessary depth is $R=25.8$.  The search efficiency is dropping in
this magnitude interval, so an exact estimate of the effective search
volume would require a model for the intrinsic KBO distance
distribution.  An approximation adequate for our purposes is to take
the effective area to be that of a $D=160$~km object at the midpoint,
60~AU.  This corresponds to a magnitude of $R=25.5$, at which the
effective search area is $0.58\,{\rm deg}^{2}$.  Our upper limit to
the surface density of KBOs with $D>160$~km at distances $55<r<65$~AU
is therefore 5~deg$^{-2}$ (95\% CL).

\section{Comparison to the Inner Kuiper Belt}

We test the hypothesis of inner-belt depletion by comparing the volume
density of KBOs inside 50~AU to that outside 50~AU.  Because we have a
faint limiting magnitude and we have determined the distances (and
hence approximate size) for all our detected objects, we can proceed
by constructing samples of objects of similar size in the two distance
regimes, {\it without any assumption about the KBO size distribution}.
The outer region is the $55<R<65$~AU annulus quantified above.

For our inner region, we select all KBOs with distances in the range
30--50~AU and diameters greater than 160~km.  Most known KBOs within
40~AU are in mean-motion resonance orbits or have likely been
perturbed by a Neptune encounter, affecting the surface density in an
uncertain manner. For this reason, we will also compare our results to
an inner region of 40--50~AU.

Our survey yields 8 KBOs which have a distance between 30--50~AU and
diameter greater than 160~km. These are all brighter than $R=24.1$,
although at 50~AU a 160~km KBO is only $R=24.7$.

Denoting the mean volume density of $D>160$~km KBOs in the inner and
outer regions as $n_1$ and $n_2$, respectively, we are interested in
an upper bound to the density ratio
\begin{math}
f \equiv n_2 / n_1.
\end{math}
We assume that $N_1$ objects were detected in the inner region over
an effective volume $V_1$, and $N_2=0$ objects were detected in the
effective volume $V_2$ of the outer region.
A Bayesian analysis with uniform prior on $n_1$ yields the following
simple formula for the probability of $f$ being above some value:
\begin{equation}
\label{bayes}
P(>f) = (1 + f V_2/V_1 )^{-N_1}
\end{equation}
The effective volume for the inner region is simply that subtended by
our survey's $1.3\,{\rm deg}^2$ effective area over the $30<R<50$~AU
distance range, since the inner-region KBOs are well above our
detection threshold.  The effective volume of the outer region is
given by the $0.58\,{\rm deg}^{2}$ effective area through the
55--65~AU depth.  The ratio of volumes between the 55--65~AU sample
and the 30--50~AU sample, $V_2/V_1$, is then
0.49. Equation~(\ref{bayes}) then gives $f<0.92$ at 95\%
confidence. If only the 7 KBOs in the inner region from 40--50~AU are
considered, then we find $f<0.67$ (95\% CL). We can thus assert in a
model-independent way that the density of KBOs in the 55--65~AU range
is at least \onethird\ {\it lower} than in the 40--50~AU range.

We could estimate the inner-region density from other surveys with
brighter magnitude limits than ours, though the comparison with our
own brighter objects has the advantage of cancelling any dependence on
ecliptic latitude or longitude which may be present in the KBO
density.  There are very few published wide-area KBO surveys to serve
as comparison samples.  \citet[JLC]{JLC} and \citet[JLT]{JLT} are the
two largest.  The JLC survey covers 8.3~deg$^2$ to a limiting
magnitude of $R=24.2$. At this depth, KBOs must have diameter
$D>200$~km, to be visible all the way to 50~AU.  Out of 15 KBOs found
by JLC, 12 have distances within 30--50~AU and $D>200$~km. Using JLC's
effective areas quoted at $R=23.2$ and $R=24.2$, a rough estimate for
the effective area for all 12 KBOs is 6.1~deg$^2$.  Our survey will
detect 200~km objects to a distance of 67~AU, if the same $R=25.5$
limit is used, so we may conservatively take the outer region in this
case to be 55--67~AU across the $0.58\,{\rm deg}^{2}$ effective area
for $R\le25.5$.  This then yields a Bayesian limit of $f<2.17$ (95\%
CL) for the ratio of outer to inner volume densities of $D>200$~km
objects.

The JLT survey, with a much brighter limiting magnitude of $R=22.5$
over 51.5~deg$^2$, detects 13 KBOs.  Of these, 12 are between
30--50~AU, but only 6 are larger than $D=320$~km, the smallest size
that can be seen at 50~AU in this survey.  These 6 KBOs are all inside
45~AU, although we will still consider the inner region
30--50~AU. Objects with $D>320$~km are detectable to 85~AU at
$R\le25.5$, so we take the outer region to be 55--85~AU.  Using the
same procedures as for the JLC survey, we calculate $f<4.9$.

We should at some point expect to find objects beyond 55~AU since the
orbits of the known scattered-disk members will carry them well beyond
this point.  The sole known KBO that is likely beyond 55~AU (1999
DG$_8$) could be a scattered-disk member; its 1-night arc is
insufficient to determine an orbit.  Seven objects with $m_R>25$ were
discovered in very deep Keck pencil-beam surveys \citep{LJ98,Ch99};
all have motions over arcs of a few hours that are consistent with
distances $<50$~AU.

Our failure to detect any scattered-disk objects beyond 50~AU does not
invalidate our argument, since we claim only a {\it fair sample} of
the population, not an exhaustive survey.  We can test the fair-sample
hypothesis by examining our detected population of $D>160$~km objects;
we claim that this population is complete to about 65~AU, and that
majority of the population is within 55~AU, {\it i.e.} that we are
capable of detecting most of the objects anywhere in their orbits.
There should consequently be no bias toward finding objects at
perihelion.  Examining the 9 objects with $D>160$~km in Table~2, we
see that 4 are closer to aphelion than perihelion (1998 KY$_{61}$,
KG$_{62}$, KR$_{65}$, and 1999 KR$_{18}$), 1 is nearer perihelion
(1998 KS$_{65}$), and 4 have uncertain $a$ (1999 JA$_{132}$,
JB$_{132}$, JD$_{132}$, and JF$_{132}$).  Our objects' distance
distribution is thus qualitatively consistent with an unbiased
sampling of their own orbits.  A quantitative comparison (KS test)
also demonstrates internal consistency.  While this is a weak test, it
is one which the full population of known KBOs fails miserably---there
is an extremely strong bias toward objects near perihelion.  Indeed
our 1999 KR$_{18}$ is the only known KBO of significant orbital
eccentricity ($e>0.1$) to be discovered near aphelion.

Our lack of detections beyond 55~AU is consistent with a density
beyond 55~AU that is at most similar to that within 50~AU, and it is
likely we have observed an under-density in the outer Kuiper Belt.
The measurements are not consistent with a large increase in the
surface density beyond 50~AU.

\section{Conclusions}

If planetary perturbations are solely responsible for the structure of
the Kuiper Belt, a dense primordial disk would be expected beyond
$\sim50$~AU where these perturbations are insignificant.  Our survey
could have detected such a disk but did not.

There are several possible explanations for this non-detection. 

\begin{enumerate}

\item
KBOs in the outer Kuiper Belt could be fainter than expected, owing to
a lower albedo or much redder color or much smaller sizes.
To explain our survey results, this would require a change in the physical
properties of KBOs beyond 50~AU.  

\item
The outer Kuiper Belt could have been dynamically excited early in the
history of the solar system by a stellar encounter \citep{Id00} most
likely during the Sun's residence in its birth cluster \citep{AL00};
by perturbations from large Neptune-scattered planetesimals
\citep{P99} or by proto-planetary cores \citep{Th99}.  Such excitation
could have increased the orbital inclinations and eccentricities of
the objects, lowering the apparent surface density on the sky and
possibly decreasing the maximum size of objects through the cessation
or reversal of the accretion process.  More extreme excitation could
have stripped away the outer Kuiper Belt.

\item
The disk is actually present, but dynamically cold \citep{Ha00}.
It could have escaped our survey by simply not intersecting our fields.
This is entirely possible, if this disk is inclined to the ecliptic plane 
by as little as 1\arcdeg. The invariable plane (the angular momentum
plane of the solar system), inclined to the ecliptic by 1.5\arcdeg, 
would be a likely candidate for the location of such a cold, dense
disk. Our Field~G is only 0\fdg6 from the invariable plane, so the
scale height of the cold disk would have to be $\ll1\arcdeg$ to have
escaped detection.  This implies a very thin, extremely dense disk.
Absence of $>50~AU$ detections in future deep survey fields would
eliminate the possibility of a cold disk.

\end{enumerate}

\acknowledgements
We thank P. Guhathakurta for his help in retrieving some of these
objects, and the staff on Cerro Tololo for their excellent support.
P. Fischer provided much of the software used to process the BTC
images.  This work is supported by NASA Planetary Astronomy grant
\#NAG5-7860; GB is further supported by grant \#AST-9624592 from the
National Science Foundation; RM is further supported by NASA Origins
of Solar systems grant \#NAG5-4300 and NASA Planetary Geophysics grant
\#NAG5-6886.

\clearpage
\begin{deluxetable}{clrrccccccc}
\rotate
\tablecolumns{11}
\tablewidth{0pt}
\tablecaption{Field Information}
\tablehead{
\colhead{Field} & \colhead{Dates of} & \colhead{$RA$} & \colhead{Dec} &
\multicolumn{2}{c}{Ecliptic Coords.} &
\colhead{Invariable} & \colhead{Exp Time} &
\colhead{Filter} & \colhead{Eff. Area} & \colhead{$m_R$} \\
 & \colhead{Observation\tablenotemark{a}} 
& \multicolumn{2}{c}{(J2000)} & \colhead{Long.} & \colhead{Lat.} &
\colhead{Lat.} & \colhead{(s)} & & \colhead{(deg$^2$)} & \colhead{Limit}
}
\startdata
A & 5/19/98             & $11^h\ 37^m\ 34^s$ &  +2 13 20  &
174.0   &  -0.19  &  -1.63  & 12$\times$480 & $R$ & \nodata  & \nodata  \\
  & 5/28/98 - 5/29/98 * & $11^h\ 37^m\ 22^s$ &  +2 14 34  &
173.9   &  -0.18  &  -1.63  & 19$\times$480 & $R$ & 0.178  & 24.9 \\

B & 5/19/98             & $13^h\ 02^m\ 36^s$ &  -6 46 00  &
197.0   &  -0.03  &  -1.61  & 18$\times$480 & $R$ & \nodata  & \nodata  \\
  & 5/28/98 - 5/29/98 * & $13^h\ 02^m\ 00^s$ &  -6 42 26  &
196.9   &  -0.09  &  -1.67  & 24$\times$480 & $R$ & 0.180  & 25.2 \\

D & 5/19/98             & $20^h\ 27^m\ 52^s$ &  -19 20 00  &
304.6   &  -0.22  &   0.24  & 18$\times$480 & $R$ & \nodata  & \nodata  \\
  & 5/28/98 - 5/29/98 * & $20^h\ 47^m\ 35^s$ &  -19 19 47  &
309.1   &  -1.40  &  0.81  & 26$\times$480 & $R$ & 0.147  & 25.4  \\

E & 5/10/99 - 5/11/99 * & $12^h\ 05^m\ 00^s$ &  -0 30 00  & 
181.4   &  0.04   &  -1.48  & 33$\times$480 & $VR$ & 0.163  & 25.8   \\
  & 5/18/99 - 5/19/99 * & $12^h\ 04^m\ 34^s$ &  -0 32 28  &
181.3   &  -0.01  &  -1.53  & 33$\times$480 & $VR$ & 0.200  & 25.3  \\

F & 5/10/99 - 5/11/99 * & $14^h\ 00^m\ 00^s$ &  -12 12 00  &
212.2   &   0.03  &  -1.50  & 30$\times$480 & $VR$ & 0.196  & 25.9 \\
  & 5/18/99 - 5/19/99 * & $13^h\ 59^m\ 21^s$ &  -12 15 15  &
212.1   &  -0.08  &  -1.61  & 33$\times$480 & $VR$ & 0.200  & 25.8 \\

G & 5/10/99 - 5/11/99   & $20^h\ 45^m\ 00^s$ &  -18 00 00  & 
308.8    &   0.05  &   0.62  & 48$\times$480 & $VR$ & \nodata   &
\nodata \\
  & 5/18/99 - 5/19/99 * & $20^h\ 45^m\ 00^s$ &  -18 00 00  &
308.8    &   0.05  &   0.62  & 44$\times$480 & $VR$ & 0.167  & 25.7 \\
\enddata
\tablenotetext{a}{Observations marked with asterisks were searched for
KBOs; unmarked observations were used only for (p)recovery.}
\end{deluxetable}

\clearpage

\begin{deluxetable}{ccccccrcc}
\tablewidth{0pt}
\tablecaption{Objects Discovered}
\tablehead{
\colhead{MPC} & \colhead{Field} & \colhead{Arc} &
\colhead{$m_R$} &
\colhead{$a$\tablenotemark{a}} & \colhead{$e$\tablenotemark{a}} & \colhead{$i$} &
\colhead{Heliocentric} & \colhead{Diameter\tablenotemark{c}} \\
\colhead{Designation} & & \colhead{Length} & &
\colhead{(AU)} & &
\colhead{(\arcdeg)} &
\colhead{Dist. (AU)} &
\colhead{(km)}
}
\startdata
1998 KD66 & B & 10 d & 24.7 & 
\nodata & \nodata & $ 6.4\pm 2.9 $ & $42.9\pm3.7$ & 117 \\
1998 KE66 & B & 10 d & 25.0 & 
\nodata & \nodata & $ 2.5\pm 0.9 $ & $41.0\pm3.4$ & 94 \\
1998 KF66 & B & 10 d & 24.5 & 
\nodata & \nodata & $ 6.7\pm 1.5 $ & $31.8\pm2.0$ & 70 \\
1998 KG66 & B & 10 d & 25.1 & 
\nodata & \nodata & $ 3.5\pm 1.5 $ & $45.2\pm4.1$ & 109 \\
1998 KY61 & D & 42 d & 23.7 & 
$44.1\pm 0.1$ & $0.05\pm0.10 $ & $ 2.1\pm 0.0 $ & $46.5\pm0.0$ & 220 \\
1998 KG62 & D & 2 opp & 22.9 &
$43.4\pm 0.0$ & $0.05\pm0.01 $ & $ 0.8\pm 0.0 $ & $45.3\pm0.0$ & 301 \\
1998 KR65 & D & 2 opp & 22.9 &
$43.5\pm 0.0$ & $0.02\pm0.00 $ & $ 1.2\pm 0.0 $ & $44.4\pm0.0$ & 289 \\
1998 KS65 & D & 2 opp & 23.7 &
$43.7\pm 0.0$ & $0.03\pm0.00 $ & $ 1.2\pm 0.0 $ & $42.3\pm0.0$ & 181 \\
1999 JV127 & E & 8 d & 23.7 & 
$18.2\pm 0.2$ & $0.15\pm0.08 $ & $19.2\pm 0.7 $ & $20.9\pm0.3$ & 43 \\
1999 JA132 & E & 9 d & 23.9 & 
$42.0\pm 3.8$ & $0.07\pm0.13 $ & $ 7.3\pm 0.7 $ & $45.2\pm1.1$ & 189 \\
E2-01\tablenotemark{b} & E & 1 d & 24.9 &
\nodata & \nodata & $7.1\pm3.5$ & $31.7\pm4.2$ & 58 \\
1999 JB132 & F & 8 d & 23.5 & 
\nodata & \nodata & $17.1\pm11. $ & $39.1\pm3.7$ & 170 \\
1999 JC132 & F & 1 d & 24.3 & 
\nodata & \nodata & $ 5.4\pm 2.1 $ & $39.0\pm2.6$ & 117 \\
1999 JD132 & F & 2 opp & 23.6 & 
$45.4\pm3.3$ & $0.22\pm0.16$ & $10.5\pm 0.0 $ & $42.8\pm0.2$ & 198 \\
1999 JE132 & F & 9 d & 24.1 & 
$32.4\pm 5.0$ & $0.20\pm0.22 $ & $29.8\pm 6.4 $ & $39.1\pm1.4$ & 129 \\
1999 JF132 & F & 9 d & 24.0 & 
\nodata & \nodata & $ 1.6\pm 0.4 $ & $43.1\pm2.1$ & 164 \\
1999 JH132 & F & 9 d & 25.5 & 
\nodata & \nodata & $ 0.6\pm 0.2 $ & $41.1\pm2.4$ & 75 \\
1999 JJ132 & F & 9 d & 25.5 & 
\nodata & \nodata & $ 3.2\pm 1.2 $ & $50.1\pm2.8$ & 111 \\
1999 JK132 & F & 9 d & 24.8 & 
\nodata & \nodata & $16.0\pm 5.9 $ & $39.0\pm2.5$ & 93 \\
1999 KT16  & F & 1 d & 25.1 & 
\nodata & \nodata & $ 8.5\pm 3.7 $ & $46.3\pm3.1$ & 114 \\
1999 KK17  & F & 9 d & 25.0 & 
\nodata & \nodata & $ 8.7\pm16. $ & $49.8\pm7.5$ & 139 \\
1999 KL17  & G & 90 d & 25.2 &
$46.2\pm 0.2$ & $0.03\pm0.17 $ & $ 2.8\pm 0.0 $ & $47.6\pm0.0$ & 115 \\
1999 KR18  & G & 89 d & 24.9 &
$43.3\pm 0.3$ & $0.21\pm0.03 $ & $ 0.6\pm 0.0 $ & $52.6\pm0.1$ & 162 \\
G3-01\tablenotemark{b}  & G & 9 d & 25.6 &
$39.9\pm2.7$  & $0.16\pm0.05$ & $1.6\pm0.1$ & $33.4\pm1.4$ & 47 \\
\enddata
\tablenotetext{a}{No data are given when the arc is too
short to provide meaningful constraint.}
\tablenotetext{b}{Objects E2-01 and G3-01 were not reported to the MPC
due to insufficient S/N on the recovery observations.}
\tablenotetext{c}{Diameters assume albedo of 0.04.}
\end{deluxetable}

\end{document}